# The directed Hausdorff distance between imprecise point sets

Christian Knauer[1], Maarten Löffler[2], Marc Scherfenberg[1], and Thomas Wolle[3]

[1] Institute of Computer Science, Freie Universität Berlin, Germany.
scherfen@mi.fu-berlin.de
[2] Dep. of Information and Computing Sciences, Utrecht University, the Netherlands.
loffler@cs.uu.nl
[3] NICTA Sydney, Australia. thomas.wolle@nicta.com.au

**Abstract.** We consider the directed Hausdorff distance between point sets in the plane, where one or both point sets consist of imprecise points. An imprecise point is modelled by a disc given by its centre and a radius. The actual position of an imprecise point may be anywhere within its disc. Due to the direction of the Hausdorff Distance and whether its tight upper or lower bound is computed there are several cases to consider. For every case we either show that the computation is NP-hard or we present an algorithm with a polynomial running time. Further we give several approximation algorithms for the hard cases and show that one of them cannot be approximated better than with factor 3, unless P=NP.

## 1 Introduction

The analysis and comparison of geometric shapes are essential tasks in various application areas within computer science, such as pattern recognition and computer vision. Beyond these fields also other disciplines evaluate the shape of objects such as cartography, molecular biology, medicine, or biometric signal processing. In many cases patterns and shapes are modeled as finite sets of points.
The *Hausdorff* distance is an important tool to measure the similarity between two sets of points (or, more generally, any two subsets of a metric space). It is defined as the largest distance from any point in one of the sets, to the closest point in the other set (see Section 1.3 for a formal definition). This distance is used extensively in pattern matching.
Data imprecision is a phenomenon that has existed as long as data is being collected. In practice, data is often sensed from the real world, and as a result has a certain error region. On the one hand, many application fields of computational geometry use algorithms that take this into account. However, these algorithms are mostly heuristics, and do not benefit from theoretical guarantees. On the other hand, algorithms from computational geometry are provably correct and efficient, often under the assumption that the input data is correct. If we want these algorithms to be used in practice, they need to take imprecision into account in the analysis. Thus not surprisingly, data imprecision in computational geometry is receiving more and more attention.

In this paper, we study several variants of the important and elementary problem of computing the Hausdorff distance under the Euclidean metric between *imprecise* point sets.

### 1.1  Related Work

The Hausdorff distance is one of the most studied similarity measures. For a survey about similarity measures and shape matching refer to [2]. A straightforward, naive algorithm computes the Hausdorff distance between two point sets $A$ and $B$ consisting of $m$ and $n$ points, respectively, in $O(mn)$ time. Using Voronoi diagrams and a more sophisticated approach the running time can be reduced to $O((m+n)\log n)$, [1].

The study of imprecision within computational geometry started around twenty years ago, when Guibas *et al.* [6] introduced *epsilon geometry* as a way to handle computational imprecision. In this model, each point is assumed to be at most $\varepsilon$ away from its given location.

For a given measure on a set of imprecise points, one of the simplest questions to ask in this model is what are the possible output values? Each input point can be anywhere in a given region, and depending on where each point is, the output will have a different value. This leads to the problem of placing the points in their regions such that this value is minimised or maximised. One of the first results of this kind is due to Goodrich and Snoeyink [5], who show how to place a set of points on a set of vertical line segments such that the points are in convex position and the area or perimeter of the convex hull is minimised in $O(n^2)$ time. A similar problem is studied by Mukhopadhyay *et al.* [10], and their result was later generalised to isothetic line segments [9].

Nagai and Tokura [11] thoroughly study the efficient computation of lower and upper bounds for a variety of region shapes and measures; in particular they study the diameter, the width, and the convex hull, and all their algorithms run in $O(n \log n)$ time. However, not all of their bounds are tight. Van Kreveld and Löffler [12] study the same problems and give algorithms to compute tight bounds, though the running times of the algorithms can be much higher and some variants are proven to be NP-hard.

### 1.2  Contribution

In this paper, we assume that an imprecise point is modelled by a disc with a given centre and radius. In general, it is possible that the discs intersect. We assume we have two sets of points, $P$ and $Q$, and that at least one of them is imprecise. We want to compute the directed Hausdorff distance from $P$ to $Q$. This includes both the tight lower and upper bound on the possible values, for each combination. This leads to six different cases. Additionally, in some settings the problems become easier if we restrict the model of imprecision to disjoint discs or discs that all have the same radius; we state these results separately. Our results are summarised in Table 1.

| setting | tight lower bound | tight upper bound |
|---|---|---|
| $h(\tilde{P}, Q)$ [general] | $O(n^2)$ | $O(n \log n)$ |
| $h(P, \tilde{Q})$ [general] | NP-hard*, 4-APX in $O(n^3 \log^2 n)$ | $O(n \log n)$ |
| [disjoint unit discs] | 3-APX-hard, 3-APX in $O(n^{10} \log n)$ | $O(n \log n)$ |
| $h(\tilde{P}, \tilde{Q})$ [general] | NP-hard | $O(n^2)$ |
| [const. depth in $\tilde{P}$] | | $O(n \log n)$ |

**Table 1.** $P$ and $Q$ are point sets and $\tilde{P}$ and $\tilde{Q}$ are imprecise point sets. Results are shown for the case when all sets have $O(n)$ elements. *can be computed exactly in $O(n^3)$ if the discs are disjoint and the answer is smaller than $r(\sqrt{5 - 2\sqrt{3}} - 1)/2$ where $r$ is the radius of the smallest disc in $\tilde{Q}$.

In the next section, we review some definitions and structures that we use to obtain our results. After that, we present our three main results. In Section 2, we give a general algorithm for computing the upper bound, which works in all settings in the table, though it can be simplified (conceptually) in some settings. In Section 3, we prove hardness of computing the lower bound in most settings. Finally, in Section 4, we give algorithmic results for computing the lower bound, exactly in some cases and approximately in others. Due to space constraints some proofs and details can be found in the appendix.

### 1.3 Preliminaries

The directed Hausdorff distance $h$ from a point set $P = \{p_1, \ldots, p_m\}$ to a point set $Q = \{q_1, \ldots, q_n\}$ with an underlying Euclidean metric can be computed in $O((n+m) \log n)$ time, see [1], and is defined as (see Fig. 1 for an example):

$$h(P, Q) = \max_{p \in P} \min_{q \in Q} \|p - q\|$$

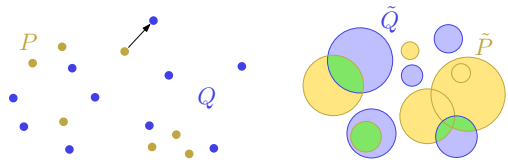

Let $\tilde{P}$ and $\tilde{Q}$ denote two imprecise point sets consisting of $m$ and $n$ closed discs respectively. We call a set $P = \{p_1, \ldots, p_m\}$ a *precise realisation* of $\tilde{P} = \{\tilde{p}_1, \ldots, \tilde{p}_m\}$ if $p_i \in \tilde{p}_i$ for all $i$. We also write $P \Subset \tilde{P}$ in this case.

**Fig. 1.** (a) $h(P, Q)$ is defined by the pair of points indicated by the arrow. (b) An example input of imprecise points.

We define the directed Hausdorff distance between a precise and an imprecise or two imprecise point sets as the interval of all possible outcomes for that distance.

$$h(P, \tilde{Q}) = \{h(P, Q) \mid Q \Subset \tilde{Q}\}, \quad h(\tilde{P}, Q) = \{h(P, Q) \mid P \Subset \tilde{P}\}$$
$$h(\tilde{P}, \tilde{Q}) = \{h(P, Q) \mid P \Subset \tilde{P}, Q \Subset \tilde{Q}\}$$

Further, we denote the tight upper and lower bounds of this interval by $h_{\max}$ and $h_{\min}$ respectively, for example

$$h_{\max}(P, \tilde{Q}) = \max h(P, \tilde{Q}) \text{ and hence } h(P, \tilde{Q}) = [h_{\min}(P, \tilde{Q}), h_{\max}(P, \tilde{Q})].$$

## 2  Algorithm for computing the tight upper bound

In this section, we consider the following problem. Given are two set of discs $\tilde{P}$ and $\tilde{Q}$. The radii may be all different; an example input is shown in Fig. 1(b). We want to place point sets $P \Subset \tilde{P}$ and $Q \Subset \tilde{Q}$ such as to maximise the directed Hausdorff distance $h(P,Q)$. In other words, we want to place the points in $P$ and $Q$ such that one point from $P$ is as far as possible away from all points in $Q$. The placements of the remaining points of $P$ do not matter. So, we need to identify which point $\hat{p} \in P$ will play this important role.

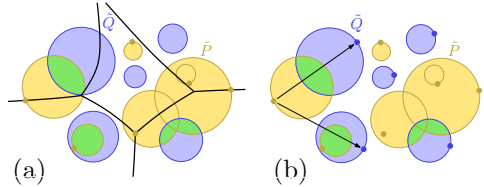

**Fig. 2.** (a) The inverted additive Voronoi Diagram (iaVD) of $\tilde{Q}$. The point set $P$ placed locally optimal. (b) The points in $Q$ are all placed as far away from $\hat{p}$ as possible.

### 2.1  Basic algorithm

We will first compute the *inverted additive Voronoi Diagram* (iaVD) of $\tilde{Q}$. This is a subdivision of the plane into regions where each point $x$ in the plane is associated with the disc in $\tilde{Q}$ whose furthest point is closest to $x$. See Fig. 2(a) for an example. This diagram can be computed in $O(n \log n)$ time [4], since it corresponds to the additively weighted Voronoi Diagram (also known as Apollonius diagram) of the centres of $\tilde{Q}$, where the weight of a point is minus the radius of the corresponding disc.

Using the iaVD, we can place each point $p \in \tilde{p} \in \tilde{P}$ at a locally optimal position, as if it were $\hat{p}$. We identify three possible placement types for $p$ that are locally optimal, as is illustrated in Fig. 2.

1. A vertex of the iaVD.
2. An intersection point between a Voronoi edge and a disc from $\tilde{P}$.
3. A point on the boundary of $\tilde{p}$ that is furthest away from the iaVD site whose cell contains the centre of $\tilde{p}$

We can now iterate over all points in $P$ and their locally optimal placements, and determine $\hat{p}$ by keeping track of the locally optimal placement $p \in \tilde{p}$ such that the shortest distance between $p$ and (the furthest point on) any disc in $\tilde{Q}$ is maximised. Once $\hat{p}$ is known, we place all points in $Q$ as far away from $\hat{p}$ as possible, and all points in $P \setminus \{\hat{p}\}$ anywhere inside their discs. The result is shown in Fig. 4(b). As it is possible that there are $O(mn)$ locally optimal placements of the second type (namely: an intersection between a disc boundary and a Voronoi edge), we conclude with the following theorem.

**Theorem 1.** *Given two sets $\tilde{P}$ and $\tilde{Q}$ of imprecise points of size $m$ and $n$, respectively, we can compute $h_{\max}(\tilde{P},\tilde{Q})$ and precise realisations $P \Subset \tilde{P}$ and $Q \Subset \tilde{Q}$ with $h(P,Q) = h_{\max}(\tilde{P},\tilde{Q})$ in $O(nm + n \log n)$ time.*

## 2.2  Faster algorithms in special cases

In this section we show how the above result can be improved under certain assumptions. To speed up the algorithm, we make some observations about the nature of locally optimal placements.

**Lemma 1.** *Let $\tilde{p}$ be a disc in $\tilde{P}$, and let $\tilde{q}_1$ and $\tilde{q}_2$ be two discs in $\tilde{Q}$, such that the part of the bisector of $\tilde{q}_1$ and $\tilde{q}_2$ that is in the iaVD slices through $\tilde{p}$ (that is, it is not connected to a vertex of the iaVD inside $\tilde{p}$). Then the optimal placement of p occurs on the same side of this bisector where the centre of $\tilde{p}$ is, regardless of what the rest of the iaVD looks like.*

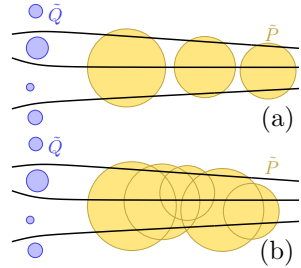

**Fig. 3.** (a) There could be a quadratic number of intersections between the edges of the iaVD of $\tilde{Q}$ and the discs in $\tilde{P}$. (b) When the discs overlap, the *union* of $\tilde{P}$ has fewer intersections with the iaVD.

*Proof.* Some notation: let $p_c$ be the centre of $\tilde{p}$, $q_{c1}$ the centre of $\tilde{q}_1$ and $q_{c2}$ the centre of $\tilde{q}_2$. Now let $f_1$ be the point on the boundary of $\tilde{p}$ that is furthest away from $q_{c1}$ (this would be the type 3 placement if $\tilde{q}_1$ was the only player), and similarly let $f_2$ be the point furthest away from $q_{c2}$. Now, suppose w.l.o.g. that $p_c$ is on the same side as $q_{c1}$. Now, suppose that the optimal placement $p$ is on the other side, that is, on the side of $q_{c2}$. Then we observe that $f_2$ must be on the side of $q_{c1}$, because $q_{c2}$, $p_c$ and $f_2$ lie on a line. This means that along the boundary of $\tilde{p}$, the intersection points with the bisector have a better value than any other point on the side of $q_{c2}$, in particular, better than $p$, which is a contradiction. (Note that if there are other cells of the iaVD involved, the value of $p$ could only be lower).

This lemma basically says that if we want to place a certain point $p$ locally optimally, we can start looking by walking from the centre of $\tilde{p}$ and never have to cross edges of the iaVD that slice through $\tilde{p}$. Like illustrated in Fig. 3. This makes us arrive at the following conclusion.

**Corollary 1.** *Let $\tilde{p}$ be a disc in $\tilde{P}$, and suppose that the iaVD has $t$ vertices inside $\tilde{p}$. Then we can find the locally optimal placement for p in $O(t)$ time.*

This immediately implies that if the discs of $\tilde{P}$ do not overlap, we can simply place all points p independently in linear time.
Now, assume that the discs of $\tilde{P}$ are disjoint, or that the intersection depth is at most some constant $c$. Then, clearly, each vertex of the iaVD can appear in at most $c$ discs of $\tilde{P}$. So, if each disc $\tilde{p}_i$ contains $t_i$ vertices of the iaVD, we have $\sum_i t_i \leq cn$, and we can find all locally optimal placements in $O(n)$ time.

**Theorem 2.** *Given two sets $\tilde{P}$ and $\tilde{Q}$ of imprecise points of size $m$ and $n$, respectively, where the discs in $\tilde{P}$ have constant intersection depth, we can compute $h_{\max}(\tilde{P}, \tilde{Q})$ and precise realisations $P \Subset \tilde{P}$ and $Q \Subset \tilde{Q}$ with $h(P,Q) = h_{\max}(\tilde{P}, \tilde{Q})$ in $O((m+n)\log(m+n))$ time.*

The algorithm described in this section works in the most general setting. However, in some more specific settings, the algorithm can be simplified. For example, when the discs of $\tilde{Q}$ are unit discs, the iaVD is simply the normal Voronoi diagram. When $P$ is not imprecise, there are of course only $m$ possible locations for $\hat{p}$, and we do not need to look for all three placement types. This results in the running times as indicated in Table 1.

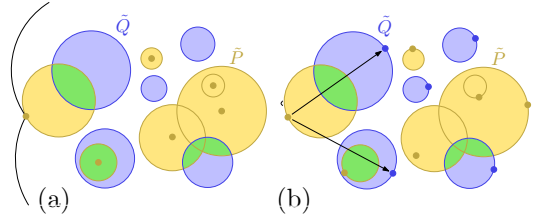

**Fig. 4.** (a) An example input. (b) The optimal output, shown as a set of circles covering $Q$.

## 3  Hardness results for tight lower bounds

In this section, we consider a transformation from the known NP-complete problem PLANAR 3-SAT [8] to the problem of computing $h_{\min}(P, \tilde{Q})$ for a set $P$ of points and a set $\tilde{Q}$ of discs with radius $r$. In the PLANAR 3-SAT problem, we are given as input a 3-SAT formula $f$ with the additional property that the graph $G(f)$ is planar, where $G(f)$ has a vertex for each variable and each clause in $f$, and there is an edge between a variable vertex and a clause vertex if the variable occurs in the clause. Having the boolean formula $f$ and a planar embedding of $G(f)$, the transformation is as follows (see Fig. 5(a,b) for a general overview):

For each variable vertex $v$ in $G(f)$, we construct a *cycle* $C$ of alternating points in $P$ and discs in $\tilde{Q}$. The distance between consecutive points and discs is $\epsilon$, such that $r = 2.5\epsilon$ (see Fig. 5(c)). There may be bends up to a certain angle, and also other geometric features necessary to connect cycles and chains. When looking only at the points $P^C$ and discs $\tilde{Q}^C$ corresponding to a cycle $C$, we observe that by the construction of $C$, there are two realisations $Q_0^C, Q_1^C \Subset \tilde{Q}^C$ such that $h(P^C, Q_0^C) = \epsilon$ and $h(P^C, Q_1^C) = \epsilon$. These two realisations represent the two possible boolean values the variable for that cycle can have.

For each edge $\{v, c\}$ in $G(f)$, we construct a *chain* of alternating points in $P$ and discs in $\tilde{Q}$ with distance $\epsilon$ (see Fig. 5(d)). The chain connects the cycle corresponding to the variable $v$ and the representation of a clause $c$. One end of this chain is a disc that will be part of a representation of clause $c$ (see Fig. 5(e)), the other end is a point $p$ that is placed near a disc $\tilde{q} \in \tilde{Q}$ of a variable cycle such that $p$ has distance $\epsilon$ to either $Q_0^C \cap \tilde{q}$ or $Q_1^C \cap \tilde{q}$ (see Fig. 5(e)).

Each clause vertex in $G(f)$ is represented by three discs and one additional point $p^*$, such that the disc centres lie on the vertices of an equilateral triangle, and the point has distance $\epsilon$ to each of the discs. The three discs are ends of chains that connect to cycles that correspond to the three literals in the clause.

**Theorem 3.** *Let $P$ be a precise point set and $\tilde{Q}$ be an imprecise point set of pairwise disjunct discs. It is NP-hard to compute a $\delta$-approximation of the directed Hausdorff distance $h_{min}(P, \tilde{Q})$ for $1 \leq \delta < 3$.*

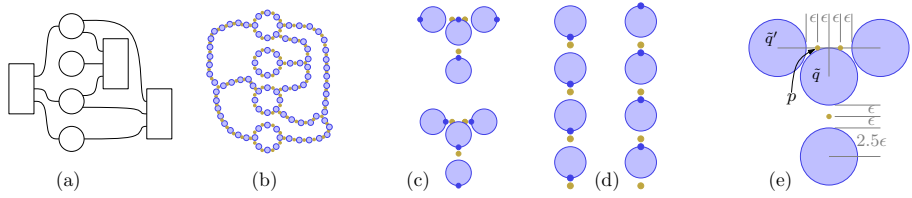

**Fig. 5.** (a) Planar embedding of $G(f)$, circles represent variables and rectangles represent clauses. (b) Rough overview of how $G(f)$ is transformed into $P$ and $\tilde{Q}$, some details are misrepresented. the chain starts with $p$ followed by $\tilde{q}'$, all other points and discs belong to the cycle. (c,d) Two realisations (representing opposite boolean values) with Hausdorff distance $\epsilon$ of chains, cycles and connections. (e) Connection of a chain to a cycle,

## 4 Algorithms for tight lower bounds

In this section we present algorithms for computing the minimum of $h(\tilde{P}, Q)$ and $h(P, \tilde{Q})$. As we have seen in the previous section, the latter problem is NP-hard and even hard to approximate in some settings. In the following we give a 4-approximation for the general case, an optimal 3-approximation for disjoint discs and an algorithm for the case which is not NP-hard when the Hausdorff distance is small. Many results in this section rely on similar ideas. Therefore, we will describe several (sub-) algorithms with different approximation factors and running times depending on the value $d$ of the optimal solution. Afterwards, we discuss how to apply them to obtain the results claimed in Table 1.

### 4.1 Algorithm PLACETOGETHER

In this section, we describe an algorithm for the case where we have an imprecise point set $\tilde{P}$ and a precise point set $Q$. We place all points in $\tilde{P}$ as close to a point in $Q$ as possible. Fig. 6(a) shows an example. For each pair $(\tilde{p}, q)$ with $\tilde{p} \in P$ and $q \in Q$ we could simply compute the placement $p \in \tilde{p}$ minimizing the Hausdorff distance and keep track of the longest distance over all pairs. This takes $O(mn)$ time. However, in practice it is probably better to compute the Voronoi diagram of $Q$ first, and locate the discs of $\tilde{P}$ in it. In the worst case, each disc could still intersect linearly many Voronoi cells (although the input needs to be contrived for this). Also, note that as soon as a disc from $\tilde{P}$ is discovered to contain a point from $Q$, we can stop the computation and just place the point there.

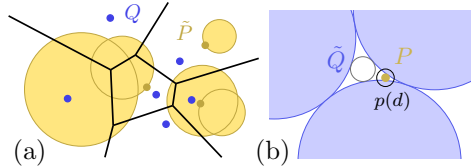

**Fig. 6.** (a) Placing points in $P$. (b) Discs of radius at most $c$ can only intersect at most two discs of $\tilde{Q}$.

**Theorem 4.** *Let $\tilde{P}$ denote an imprecise point set consisting of m discs and Q denote a precise point set consisting of n points. The tight lower bound of $h(P,\tilde{Q})$ can be computed in $O(mn)$ time.*

### 4.2   Subalgorithm CANDIDATES

In the case where $P$ is precise and $\tilde{Q}$ is imprecise, we start with a simple subalgorithm CANDIDATES to establish the following lemma. The algorithm proving this lemma can be found in Appendix B. The result will be used later.

**Lemma 2.** *Let $P$ denote a precise point set consisting of m points and $\tilde{Q}$ denote an imprecise point set consisting of n discs. It is possible to reduce the possible values of $h_{min}(P,\tilde{Q})$ to $O(m^3 + m^2n)$ many candidates in $O(m^3 + m^2n)$ time.*

### 4.3   Algorithm INDEPENDENTSETS

This algorithm computes exactly the Hausdorff distance from a precise point set $P$ to an imprecise point set $\tilde{Q}$ when the distance is small. This is an exception to the general NP-hardness of that setting.

First we compute the set of possible candidates for $h_{\min}(P,\tilde{Q})$ by CANDIDATES and discard all values greater or equal than $c = r(\sqrt{5 - 2\sqrt{3}} - 1)/2$. Now we perform a binary search on the remaining values in order to determine the smallest value $d$ for which the predicate described below evaluates to true. If such a candidate exists, the algorithm returns $d$ as the value of the bound. Otherwise $h_{\min}(P,\tilde{Q}) \geq r(\sqrt{5 - 2\sqrt{3}} - 1)/2$.

Let $p(d)$ denote the disc of radius $d$ around a point $p \in P$. There must be at least one point of $Q$ in $p(d)$ to which $p$ can be matched within a distance smaller or equal than $d$. The computation of the predicate relies on two observations: All considered values are so small that no $p(d)$ intersects more than two discs of $\tilde{Q}$. Note that a disc that intersects more than two disjoint discs of $\tilde{Q}$ has a radius of at least $r(2/\sqrt{3} - 1)$, which is greater than $c$, see Fig. 6(b). Thus, there are at most two possible matching partners for each point $p \in P$.

The second observation is that each $p(d)$ has to intersect at least one disc of $\tilde{Q}$, otherwise the Hausdorff distance would be greater than $d$ at $p$.

We define a point $p \in P$ to have degree 1 if $p(d)$ intersects just one disc $\tilde{q} \in \tilde{Q}$ and to have degree 2 if it intersects two discs of $\tilde{Q}$.

The predicate tests, whether $h_{\min}(P,\tilde{Q}) \leq d$. To this end we associate with each $\tilde{q}_i \in \tilde{Q}$ a *feasible region* $F_i$ and a set $C_i$ called *children* of $\tilde{q}_i$. The feasible region contains the valid placements of $q_i \in \tilde{q}_i$. The children of a point $\tilde{q}_i$ are the points of $P$ that can only be matched to $\tilde{q}_i$ because otherwise $h_{\min}(P, \tilde{Q})$ would be greater than $d$. In other words, the children of $\tilde{q}_i$ demand that $q_i$ is placed in its feasible region $F_i$, see Fig. 7(a).

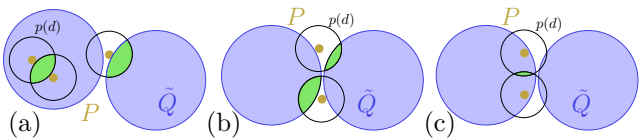

**Fig. 7.** A point $q \in \tilde{q}$ may only be placed within its feasible region (green). (a) The two left points of $P$ have degree 1, the right point has degree 2. (b,c) The points have degree two. Both cases show a scenario which allows to match the points locally, i.e. only considering the set $D$ and its two corresponding feasible regions.

We restrict the feasible regions and children in an iterative manner with the help of the sub-algorithms REMOVE-DEGREE-1-DISCS and REMOVE-DEGREE-2-DISCS, which can be found in C.1. If a feasible region turns out to be empty, the computation stops and the predicate returns false. The first sub-algorithm considers only points $p$ of degree 1 and restricts the feasible regions of the discs intersected by $p(d)$. The second sub-algorithm computes which $p(d)$ of degree 2 can also only by stabbed by one disc in $\tilde{Q}$ by considering all $p(d)$ which intersect the same two feasible regions, see Fig. 7(b) and 7(c). Afterwards there are only points $p$ unmatched whose disc $p(d)$ can be stabbed in two valid feasible regions. Furthermore, two discs $p(d)$ cannot intersect if they belong to two different pairs of feasible regions, because we only consider distances $d < r(\sqrt{5 - 2\sqrt{3}} - 1)/2$. Thus, it is possible to check for a valid point matching of the remaining points by computing the maximum matching in a bipartite graph (see section C.1 for a detailed description). Finally, we make use of a maximum matching in order to check, whether all points can be matched within the distance $d$.

It is simple to return a matching which realises the Hausdorff distance which the predicate proved to be realisable. Therefore, we first consider the feasible regions which are adjacent to a vertex in the maximum matching. We place the point in such a feasible region such that it intersects all discs in the adjacent set $D$ of discs. For all other $\tilde{q}_i \in \tilde{Q}$ we place their point $q_i$ somewhere within its feasible region $F_i$.

**Theorem 5.** *Let $P$ denote a precise point set consisting of $m$ points and $\tilde{Q}$ denote an imprecise point set consisting of $n$ disjoint discs. Algorithm* INDEPENDENTSETS *computes whether the tight lower bound for $h(P, \tilde{Q})$ is smaller than $r(\sqrt{5 - 2\sqrt{3}} - 1)/2$ where $r$ is the radius of the smallest disc in $\tilde{Q}$. If this is the case, the exact value of $h_{min}(P, \tilde{Q})$ is computed. The running time is $O(m^3 + m^2 n + n \log^2 n)$.*

### 4.4 Algorithm GROWNDISCS

In this section, we present an approximation algorithm for precise $P$ and imprecise $\tilde{Q}$. As a subroutine in this algorithm, we assume that we are given an algorithm that computes a $c$-approximation to the geometric $k$-centre problem (see Section 1.3), in time $T(k, n)$. We need this because when we have $k$ discs of $\tilde{Q}$ which partially overlap, and there are $n$ points of $P$ in the overlap, computing a lower bound on the Hausdorff distance for this subset is exactly solving the geometric $k$-centre problem. Using this subroutine, we will show how to get a $(c+2)$-approximation to our problem in time $O(m^3 + m^2n + n^2T(k,m)\log(m+n))$. Fig. 8 shows an example of the problem. We first compute the set of possible values of the Hausdorff distance using Algorithm CANDIDATES , followed by a binary search on the resulting candidate values in order to determine the smallest value $d$ for which the predicate described below evaluates to true. For any value $d$, the decision will return a solution with distance at most $(c+2)d$ if a solution of value $d$ exists. Assume that this is the case. We grow the discs in $\tilde{Q}$ by $d$, and consider the resulting arrangement of enlarged discs $\mathcal{A}$. Fig. 9(a) shows an example of this. We observe that all points of $P$ need to be inside some cell of this arrangement, otherwise there exists no solution of distance $d$. Now each cell of the arrangement contains a subset of the points from $P$, which are covered by a number of circles of radius $d$, see Fig. 9(b). Now we can compute an approximate solution independently in each cell of $\mathcal{A}$ by applying the approximation algorithm for geometric $k$-centre. This provides us with a number of circles per cell of the arrangement. Each cell can also be identified with a subset of the discs of $\tilde{Q}$ whose enlarged discs contain this cell. To solve the problem, we need to find a matching between the discs of $\tilde{Q}$ and the circles that cover $P$. In Appendix D we describe more details of the decision algorithm, which runs in $O(n^2T(k,m) + mn + m\sqrt{m})$ time. For the total running time, we first spend $O(m^3 + m^2n)$ time to execute Algorithm CANDIDATES and compute the possible values of $d$. So, the total time we spend is $O(m^3 + m^2n + (n^2T(k,m) + mn + m\sqrt{m})\log(m+n))$.

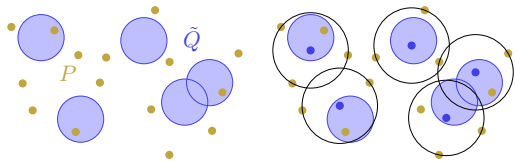

**Fig. 8.** (a) A set $P$ of precise points and a set $\tilde{Q}$ of imprecise points. (b) The optimal output. A set of circles of radius $d$ is shown around the points in $Q$, which cover the points in $P$.

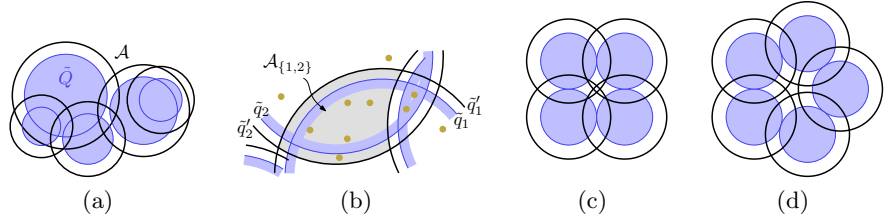

(a)             (b)             (c)             (d)

**Fig. 9.** (a) The arrangement $\mathcal{A}$ formed by the expanded discs $\tilde{Q}'$. (b) Each cell of the arrangement is determined by the indices of the discs it lies inside. (c,d) Four disjoint discs of radius $r$ of $\tilde{Q}$ such that the arrangement of enlarged circles of radius $1\frac{1}{2}r$ have a common intersection, while with five discs, this is not possible anymore.

**Theorem 6.** *Let $P$ denote a precise point set consisting of $m$ points and $\tilde{Q}$ denote an imprecise point set consisting of $n$ discs. Given a $c$-approximation to the geometric $k$-covering problem that runs in $T(k,m)$ time, we can compute a $(c+2)$-approximation to the tight lower bound of $h(P,\tilde{Q})$ in $O(m^3 + m^2n + (n^2T(k,m)+mn+m\sqrt{m})\log(m+n))$ time, where $k \leq n$ is an internal parameter of the optimal solution.*

### 4.5 Putting the algorithms together

For the remainder of this section let $r_{\min}$ and $r_{\max}$ denote the radius of the smallest and largest disc in $\tilde{Q}$. When $P$ is precise and $\tilde{Q}$ is imprecise, we note that by Theorem 6 Algorithm GROWNDISCS immediately presents a 4-approximation for the case when the discs may have different radii and overlap, which we obtain by plugging in a 2-approximation algorithm for geometric $k$-covering that runs in $O(m \log k)$ time [3]. The running time of the entire algorithm then becomes $O(m^3 + m^2 n + mn^2 \log(m+n) \log n)$ in the worst case.

We can improve this algorithm by first testing whether $v < c \cdot r_{\min}$ using Algorithm INDEPENDENTSETS and Theorem 5, without increasing the asymptotic running time. If it is, then we can actually compute the exact solution.

Furthermore, when the discs are disjoint and all have the same size, we can improve this result to a 3-approximation by combining Algorithm GROWNDISCS and a trivial algorithm called CENTREPOINTS which simply places every imprecise point at the centre of its disc. First we test whether $v > r/2 = r_{\max}/2$, by applying CENTREPOINTS and checking whether the resulting Hausdorff distance is larger than $3/2r$. If it is, we are done. Otherwise, note that each cell of $\mathcal{A}$ is a subset of the intersection of $k \leq 4$ discs, because $\tilde{Q}$'s discs are disjoint and $v < r/2$, see Fig. 9(c) and 9(d). Therefore, by Theorem 6 we can obtain a 3-approximation from Algorithm GROWNDISCS by plugging in an exact algorithm to solve the geometric $k$-covering problem.

We can solve the geometric 4-covering problem exactly by computing the arrangement circles around the points to be covered or radius $d$. The arrangement has quadratic complexity. Then we need to find out whether there are three cells that are together in all cells. There are $O(m^8)$ such combinations to test, and by keeping track of which discs are already taken care of each can be tested in constant time. So, using this algorithm, we have a $1 + 2 = 3$-approximation to the original problem for disjoint unit discs. The total running time now becomes $O(n^2 m^8 \log(m+n))$.

**Theorem 7.** *Let $P$ denote a precise point set consisting of $m$ points and $\tilde{Q}$ denote an imprecise point set consisting of $n$ disjoint discs of the same radius. The tight lower bound for $h(P, \tilde{Q})$ is 3-approximable in time $O(m^3 + m^2 n + n \log^2 n)$.*

## 5 Conclusions and Future Work

We studied computing tight lower and upper bounds on the directed Hausdorff distance between two point set, when at least one of the sets has imprecision. We gave efficient exact algorithms for computing the upper bound, prove that computing the lower bound is NP-hard in most settings, and provide approximation algorithms. Furthermore, we show that in one special case, our approximation algorithm is optimal. In other settings, a gap in the factor between the hardness result and approximation still remains. When both sets are imprecise, we don't have any constructive results for the lower bound.

All our results hold for the directed Hausdorff distance. An obvious next step would be to extend them to the undirected Hausdorff distance. We can immediately solve the upper bound problem in that case using our results, since it is just the minimum of the two directed distances. However, computing lower bounds seems to be more complicated, because there one needs to find a single placement of both point sets that minimises the distance in both directions at the same time.

Other directions of future work include looking at other underlying metrics than the Euclidean metric, other similarity measures than the Hausdorff distance, or, as is common in shape matching, allowing some transformation of the point sets.

# References

1. H. Alt, B. Behrends, and J. Blömer. Approximate matching of polygonal shapes. *Ann. Math. Artif. Intell.*, 13:251–266, 1995.
2. H. Alt and L. Guibas. *Handbook on Computational Geometry*, chapter Discrete Geometric Shapes: Matching, Interpolation, and Approximation - A Survey, pages 251–265. 1995.
3. T. Feder and D. H. Greene. Optimal algorithms for approximate clustering. In *Proc. 20th Ann. ACM Symp. on Theory of Comp.*, pages 434–444, 1988.
4. S. J. Fortune. A sweepline algorithm for Voronoi diagrams. *Algorithmica*, 2:153–174, 1987.
5. M. T. Goodrich and J. Snoeyink. Stabbing parallel segments with a convex polygon. *Comput. Vision Graph. Image Process.*, 49:152–170, 1990.
6. L. J. Guibas, D. Salesin, and J. Stolfi. Epsilon geometry: building robust algorithms from imprecise computations. In *Proc. 5th Annu. ACM Sympos. Comput. Geom.*, pages 208–217, 1989.
7. J. E. Hopcroft and R. M. Karp. An $n^{\frac{5}{2}}$ algorithm for maximum matching in bipartite graphs. *SIAM Journal on Computing*, 4:225–231, 1973.
8. D. Lichtenstein. Planar formulae and their uses. *SIAM J. Comput.*, 11:329–343, 1982.
9. A. Mukhopadhyay, E. Greene, and S. V. Rao. On intersecting a set of isothetic line segments with a convex polygon of minimum area. In *Proc. 2007 International Conference on Computational Science and Its Applications*, volume 4705 of *LNCS*, pages 41–54, 2007.
10. A. Mukhopadhyay, C. Kumar, E. Greene, and B. Bhattacharya. On intersecting a set of parallel line segments with a convex polygon of minimum area. *Inf. Proc. Let.*, 105(2):58–64, 2008.
11. T. Nagai and N. Tokura. Tight Error Bounds of Geometric Problems on Convex Objects with Imprecise Coordinates. In *Japanese Conference on Discrete and Computational Geometry*, pages 252–263, 2000.
12. M. van Kreveld and M. Löffler. Largest bounding box, smallest diameter, and related problems on imprecise points. In *Proc. 10th Workshop on Algorithms and Data Structures*, LNCS 4619, pages 447–458, 2007.

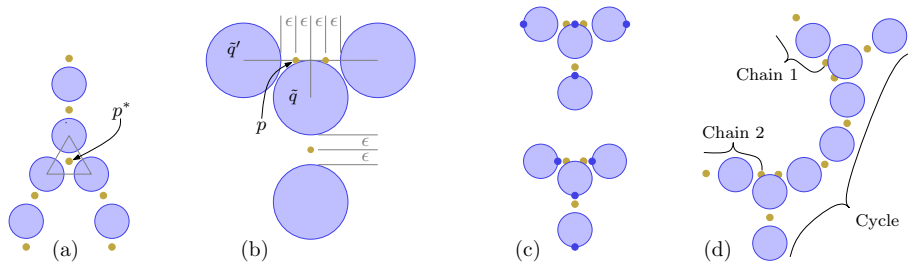

**Fig. 10.** (a) the endings of three chains arranged to representing a clause; (b) connection of a chain to a cycle with geometric details, the chain starts with $p$ followed by $\tilde{q}'$, all other points and discs belong to the cycle; (c) two realisations with Hausdorff distance $\epsilon$ of the structure in the left subfigure; (d) two chains connecting to the same cycle where the chains tap opposite boolean values;

## A  Proof of Theorem 3

Fig. 10 gives additional details necessary for the proof and the transformation.

*Proof (of Theorem 3).* For a given instance $f$ to the PLANAR 3-SAT problem, let $G(f)$ be the planar graph corresponding to $f$, embedded such that all variables are on a line, and all clauses are on either side of them, see Fig 5(a) ($G(f)$ can always be drawn in this way [8]). From this embedding, we compute (as described in Section 3) a set $P$ of precise points, a set $\tilde{Q}$ of imprecise points, and numbers $\epsilon > 0$ and $r = 2.5\epsilon$.

*Claim (1).* If $f$ is satisfiable, then $h_{\min}(P, \tilde{Q}) = \epsilon$.

*Proof.* We consider an assignment with boolean values of the variables in $f$, such that $f$ is satisfied, and we need to show that there exists a realisation $Q \Subset \tilde{Q}$, such that $h(P, Q) \leq \epsilon$. (Note that by construction, there is no realisation $Q' \Subset \tilde{Q}$, such that $h(P, Q') < \epsilon$.) For each cycle $C$ of a variable, we choose either $Q_0^C$ or $Q_1^C$ as realisations of the imprecise point set $\tilde{Q}^C$, depending on whether the variable is false or true. Discs on chains are realised in the following way: at the ending of the chain that connects to a cycle $C$, we have a point $p$ near a disc $\tilde{q} \in \tilde{Q}^C$, and $\tilde{q}$ is realised by a point $q$. The next object along the chain is a disc $\tilde{q}'$ (see Fig. 10(b)). We realise $\tilde{q}'$ in either of two ways as depicted in Fig. 10(c), depending on whether the distance between $p$ and $q$ is equal to $\epsilon$ or greater than $\epsilon$. This corresponds to a variable being either true or false. And the boolean value of the corresponding literal is then propagated to the other end of the chain to a clause $c$. Since $f$ is satisfiable, there is at least one literal in each clause that satisfies the clause. Hence, there is at least one chain with a realisation such that the point $p^*$ has distance at most $\epsilon$ to a point of this realisation. □

*Claim (2).* If $h_{\min}(P, \tilde{Q}) < 3\epsilon$, then $f$ is satisfiable.

*Proof (of Claim (2)).* We consider a realisation $Q \Subset \tilde{Q}$ with $h(P, Q) < 3\epsilon$, and we need to construct a variable assignment that satisfies $f$. We first observe that the only way where two points in $P$ can be matched to the same point $q \in \tilde{q} \in \tilde{Q}$ is where a chain connects to a cycle. (Otherwise, the distance between the two points in $P$ is larger than $6\epsilon$, and hence, they cannot be matched to the same point $q$.) And in this case, one of the points in $P$ is the end point of a chain, the other point in $P$ belongs to a cycle, and $\tilde{q}$ belongs to the same cycle (see Fig. 10(b)). From this we make an observation about how the points along chains and cycles are matched to discs along the same chains and cycles. Let us consider the sequence $p_0, \tilde{q}_0, p_1, \tilde{q}_1, p_2, \tilde{q}_2, ...$ of points and discs ordered along a fixed cycle $C$. Exactly one of the following two things is true for all $i = 0, 1, 2, ...$ (modulo length of $C$):

- $p_i$ is matched to a point $q_i \in \tilde{q}_i$, i.e. $||p_i, q_i|| < 3\epsilon$; or
- $p_i$ is matched to a point $q_{i-1} \in \tilde{q}_{i-1}$, i.e. $||p_i, q_{i-1}|| < 3\epsilon$

In other words, each point on $C$ is matched to the next disc on $C$ in clockwise order, or each point on $C$ is matched to the next disc on $C$ in counter-clockwise order, but there is no mix of these along $C$. From these two possibilities for cycle $C$, we derive the boolean value of the variable corresponding to $C$. This assignment is in accordance with the two realisations $Q_0^C$ and $Q_1^C$ (as defined above), which represent false and true. What is left to show is that this assignment satisfies $f$. To see this, we consider any clause $c$ of $f$ and argue that $c$ is satisfied. From the construction, we know that $c$ is represented by one point $p^* \in P$ and three discs being the endings of three chains. There must be a point $q \in Q$ such that $||p^*, q|| < 3\epsilon$, and $q$ must lie in one of the discs that represent the clause $c$. This disc $\tilde{q}_0$ is the ending of a chain $\tilde{q}_0, p_0, \tilde{q}_1, p_1, \tilde{q}_2, ..., p_j$. In a similar way as above, we conclude for this chain that:

- $p^*$ is matched to a point $q_0 \in \tilde{q}_0$, i.e. $||p^*, q_0|| < 3\epsilon$; and
- $p_i$ is matched to a point $q_{i+1} \in \tilde{q}_{i+1}$, for $i = 0, ..., j-1$; and
- $p_j$ is matched to a point $q_j \in \tilde{q}_j$, for some disc $\tilde{q}_j$ on some cycle $C$

The variable corresponding to $C$ has a boolean value, according to the realisation of the discs along $C$. Depending on whether this variable occurs negated or non-negated in the clause $c$, the chain $\tilde{q}_0, p_0, \tilde{q}_1, p_1, \tilde{q}_2, ..., p_j$ is connected to the cycle $C$, such that "the boolean value true is propagated along the chain". Hence, by construction we have that the boolean value of the variable corresponding to $C$ satisfies the clause $c$.
□

We conclude the proof of the theorem by observing that the construction can be done without any intersection between discs and/or points, and such that chains and/or cycles are far enough apart from each other not to interfere. We also note that the size of $P$ and $\tilde{Q}$ is polynomial in the size of $G(f)$, which follows from our planar embedding of $G(f)$.
□

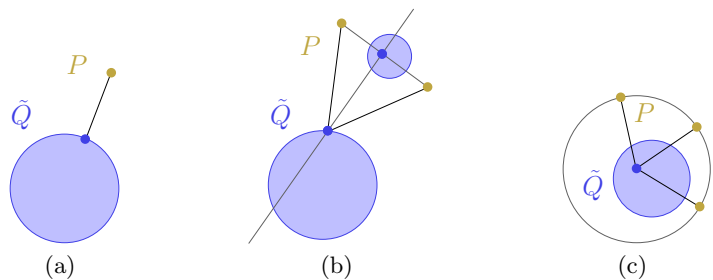

**Fig. 11.** There are only polynomially many candidates for the infimum of $h(P, \tilde{Q})$ which are determined by (a) one point of $P$, (b) by two points of $P$ or (c) three or more points of $P$.

## B  Subalgorithm CANDIDATES

Let $q \in \tilde{q} \in \tilde{Q}$ be a placement of an imprecise point in $\tilde{Q}$ which realises the Hausdorff distance $d$. The distance $d$ can be determined by $q$ together with one, two, or three points of $P$. If $d$ is determined by one point of $P$ there are $O(mn)$ possibilities, see Fig. 11(a). If $d$ is determined by two points $p_1, p_2 \in P$ the point $q$ lies on the bisector of the line segment $p_1 p_2$, see Fig. 11(b), for which $O(nm^2)$ possibilities exist. Finally, if $d$ is determined by three (or more) points all these points lie on a circle whose centre is $q$. Thus there are $O(m^3)$ possible locations, see Fig. 11(c). The algorithm simply computes and returns all $O(m^3 + m^2 n)$ locations in $O(m^3 + m^2 n)$ time.

## C  Additions to INDEPENDENTSETS

### C.1  Subalgorithms

The following two sub-algorithms restrict the feasible region and children of each disc $\tilde{q} \in \tilde{Q}$ in an iterative manner.

Before calling REMOVE-DEGREE-1-DISCS we define a set $P_R$ of all points which are not matched so far and set $P_R := P$.

REMOVE-DEGREE-1-DISCS
  **1 forall** $\tilde{q}_i \in \tilde{Q}$ **do**
  **2**     set $F_i := \tilde{q}_i$
  **3**     set $C_i := \varnothing$
  **4 while** *there is some $p \in P$ such that $p(d)$ intersects only one $F_i$* **do**
  **5**     set $F_i := F_i \cap p(d)$
  **6**     **if** $F_i = \varnothing$ **then**
  **7**         **return** false
  **8**     set $C_i := C_i \cup \{p\}$
  **9** set $P_R := P_R \setminus \bigcup_i C_i$
**10** REMOVE-DEGREE-2-DISCS

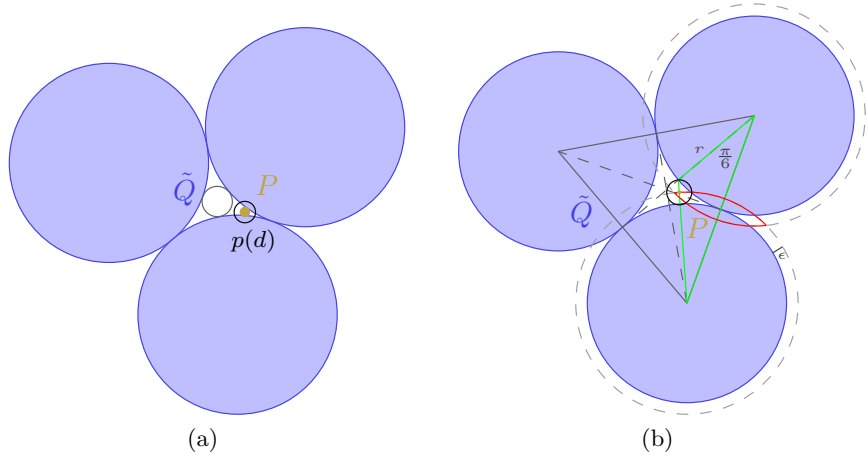

**Fig. 12.** (a) The gray circle in the middle has the minimal radius of $r(2/\sqrt{3} - 1)$ which is necessary to intersect at least three discs of $\tilde{Q}$, where $r$ denotes the radius of the blue discs. Because all considered candidates for the minimal Hausdorff distance are smaller than the radius of the gray disc, there are at most two possible matching partners for each point in $P$. (b) The black circle has the maximal radius $c$ for which no two circles intersecting two different pairs of discs in $\tilde{Q}$ can be stabbed by only one point $q \in \tilde{q}$. Since it has to intersect two discs in $\tilde{Q}$, its centre must lie within the red lune of these discs grown by $c$. Furthermore, its boundary must not intersect the green segment denoted by $r$ because it could intersect another black circle intersecting the upper two discs of $\tilde{Q}$, otherwise. Using the cosine formula it holds that $(r + 2c)^2 < r^2 + (2r)^2 - 2r2r\cos\frac{\pi}{6}$. Solving the latter inequality for $c$ yields $c < r(\sqrt{5 - 2\sqrt{3}} - 1)/2$.

In line 9 the remaining points $P_R = P \setminus \bigcup_i C_i$ are points whose disc $p(d)$ intersects exactly two feasible regions. It is still possible to match points in $P$ to points in $\tilde{Q}$ by only analysing their local environment. This is done by REMOVE-2-DISCS.

REMOVE-DEGREE-2-DISCS
1  **foreach** *pair of feasible regions* $(F_i, F_j), i \neq j$ **do**
2      compute the set $D$ of discs $p(d)$ intersecting both $F_i$ and $F_j$
3      **if**
4      $D$ can be stabbed by one point of either $F_i$ or $F_j$   or
5      $D$ needs one point of $F_i$ and one point of $F_2$ to be stabbed
6      **then**
7          restrict $F_i$ and $F_j$ accordingly
8          **if** $F_i = \emptyset \lor F_j = \emptyset$ **then**
9              **return** false
10         REMOVE-DEGREE-1-DISCS
11 BUILDGRAPH

Note, that all sets $D$ of line 2 partition the set of the discs $p(d)$ of the points in $P_R$. Line 7 restricts the matching for points of degree 2 whose matching does not interfere with the matching of points with other pairs of feasible regions. See Fig. 7(b) and 7(c) for an illustration of the two possible scenarios allowing a local matching.

In line 11 all discs of each subset $D$ can be stabbed by only one point of the two feasible regions $F_i$ and $F_j$ they intersect. Further, it holds that no two discs $p(d)$ of different sets $D$ can be stabbed by only one point, because $d < r(\sqrt{5 - 2\sqrt{3}} - 1)/2$, see Fig. 12(b).

Thus, it is possible to check for a valid point matching of the remaining points in $P_R$ by computing the maximum matching in a bipartite graph as follows. BUILDGRAPH builds a bipartite graph on the feasible regions and the sets $D$ of the partition of the discs $p(d)$ of the points in $P_R$. For each cell $D$ of the partition there are two edges in the graph connecting $D$ with the two feasible regions that the discs in $D$ intersect. We now compute a maximum matching on that graph. If this matching connects all $D$-vertices with a feasible region, the predicate returns true and the bound for the Hausdorff distance is smaller or equal than $v$. Otherwise the predicate returns false.

### C.2 Running time

The algorithm consists of three phases: It first computes a polynomial set of candidates which takes $O(m^3 + m^2 n)$ time. On this set we perform a binary search using the predicate. The computation of the predicate is done by some recursive calls of the two sub-algorithms REMOVE-DEGREE-1-DISCS and REMOVE-DEGREE-2-DISCS. These need to know the intersections of the discs $p(d)$ with the feasible regions, which are the discs in $\tilde{Q}$ in the beginning. We store these intersections distributed with every point $p \in P$ and store references with each feasible region to the $p(d)$ it intersects. The initial set of the intersections can be computed using a sweep-line in $O((m+n) \log(m+n))$ time. The restrictions of the feasible regions can take at most $O(m)$ time. Further we maintain one point set containing points $p \in P$ with degree 1 and a second point set for points of degree 2. We move a point from the second to the first point set if its degree is decreased. Thus, having the initial intersection set, all calls of REMOVE-DEGREE-1-DISCS without line 10 take $O(m)$ time.

The sub-algorithm REMOVE-DEGREE-2-DISCS needs to iterate over all pairs of feasible regions. Instead of considering all possible pairs we only maintain a set of region pairs which indeed intersect some discs $p(d)$. Because all $D$'s partition the points in $P$ there are at most $m$ discs to consider in the stabbing analysis from line 1 to 6, thus REMOVE-DEGREE-2-DISCS needs $O(m)$ time per call. Since it is called at most $m$ times by REMOVE-DEGREE-1-DISCS its overall running time is $O(m^2)$.

Finally we need to compute a maximum matching in the bipartite graph. Using the algorithm of Hopcroft and Karp [7] this needs $O(m\sqrt{m+n^2})$ time.

Putting all together we get a running time of $O(m^3 + m^2n + ((m+n)\log(m+n) + m^2)\log(m^3 + m^2n) + m\sqrt{m+n^2})$ which can be simplified to $O(m^3 + m^2n + n\log^2 n)$.

## D   Decision algorithm used in GrownDiscs

Let $d$ be any given positive value. We will describe a decision algorithm that returns, if there exists a solution to our problem with distance at most $d$, a solution with distance at most $(c+2)d$. If no solution of distance at most $d$ exists, the algorithm either still returns a solution with distance at most $(c+2)d$, or it returns false.

Let $\tilde{q}_1, \ldots, \tilde{q}_n$ be the discs in $\tilde{Q}$ where disc $\tilde{q}_i$ has radius $r_i$. We define the *grown* disc $\tilde{q}'_i$ to be the disc with the same centre point as $\tilde{q}_i$, but with radius $r_i + d$. We call the resulting set $\tilde{Q}'$.

**Observation 1** *If $P$ is not covered by $\tilde{Q}'$, there exists no solution of value $d$.*

So, we assume $P$ is covered by $\tilde{Q}'$. We can test this easily, and immediately return false if this is not the case. Now, we compute the arrangement $\mathcal{A}$ of $\tilde{Q}'$, which has quadratic complexity in the worst case. Fig. 9(a) shows an example of the arrangement formed by the discs $\tilde{Q}'$. If $I \subseteq \{1, \ldots, n\}$ is a certain set of indices, denote by $\mathcal{A}_I$ the cell of the arrangement in the intersection of all discs $\{\tilde{q}'_i \mid i \in I\}$, but not inside any other disc. (Of course, most of these cells do not exist, since there is only a quadratic number of cells.) Each cell of this arrangement contains a subset of $P$; we define $P_I$ to be the set of points of $P$ inside $\mathcal{A}_I$. Fig. 9(b) shows an example.

Now, assume that there exists a solution of value at most $d$.

**Observation 2** *Let $I$ be a set of indices. In the optimal solution $Q \in \tilde{Q}$, all the points of $P_I$ are covered by circles of radius $d$ around the points in $\{q_i \mid i \in I\}$.*

*Proof.* Since the optimal solution has Hausdorff distance $h(P, Q) \leq d$, we know that each point $p \in P$ is covered by some circle of radius $d$ around a point $q \in Q$. Now assume that $p \in P_I$. Then we know that $|pq| \leq d$, and $q \in \tilde{q}$, therefore $p \in \tilde{q}'$. So, by definition of $\mathcal{A}_I$, $q$ must be $q_i$ for some $i \in I$.  □

This observation suggests we can solve the problem somehow separately in each cell of $\mathcal{A}$. For a given cell $\mathcal{A}_I$, the optimal solution uses $k_I \leq |I|$ circles of radius $d$ (centred around points of $Q$) to cover the points in $P_I$. Now, we could compute such a set of circles (most likely a different set) by applying the c-approximation algorithm for geometric $k$-centre. This would provides us with a set $C_I$ of $k'_I \leq k_I$ circles of radius $cd$. However, there is a problem with this approach. The solutions are not independent: it is possible that a certain circle of the optimal solution covers points from two different cells of the arrangement. This means we may have constructed more than $n$ circles.

So, what we do instead is this. We process the cells of the arrangement in any order. For the first cell $\mathcal{A}_I$, we compute a set $C_I$ of at most $k_I$ circles of radius $cd$ that cover $P_I$. Now, we grow our circles until they have radius $(c+2)d$. This ensures that any points of $P$ outside $\mathcal{A}_I$ that were covered by discs of the optimal solution that were covering any points of $P_I$, are now also covered by $C_I$. Fig. 13 illustrates this.

A second complication comes from the fact that we required the centres of the circles to be in $\tilde{Q}$, not just in $\tilde{Q}'$. In order to ensure this, we simply move the circle centres to the closest point in their discs, moving them by at most $d$. Since the circles are enlarged by $2d$, the moved and enlarged circles will still cover all points of $P$ that were covered by the original circles. Fig. 14 shows this case. Furthermore, this case does not interfere with the case described above, because a circle cannot at the same time be close to the boundary of $\mathcal{A}_I$ and far enough away from it not to cover a point that is covered by a circle that also covers a point from a neighbouring cell.

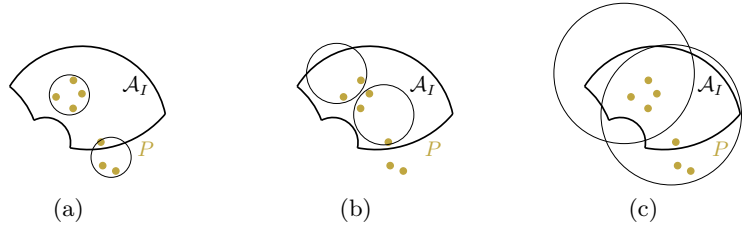

**Fig. 13.** (a) A cell $\mathcal{A}_I$ of the arrangement, and a set of two circles of radius $d$ covering $P_I$ in the optimal solution. (b) A different set of at most two circles of radius $cd$ covering $P_I$, as produced by the subroutine. (c) The enlarged circles with radius $(c+2)d$ also cover all points outside $\mathcal{A}_I$ that could be covered by the circles of the optimal solution.

For each next cell, we only consider those points that have not been covered yet, and otherwise proceed in the same way.

This procedure results in a set $C$ of at most $n$ circles, composed of a set $C_I$ for each cell $\mathcal{A}_I$ of the arrangement. This set has the property that each $C_I$ contains no more circles than the corresponding set in the optimal solution. This implies that there exists a matching between $C$ and $\tilde{Q}'$ in the graph that has an edge between circle $c$ and disc $\tilde{q}'_i$ if $c$ is in a set $C_I$ where $i \in I$. Clearly, this means that the centre of $c$ is inside $\tilde{q}'_i$. Since an optimal matching exists, we can also compute one efficiently (although it may be a different one).

For each value $d$, we spend $O(n^2)$ time to compute $\mathcal{A}$, and $T(k_I, |P_I|)$ time per cell to solve the geometric $k$-centre problem. If $k$ is the largest value of $k_I$ over all $I$, then a crude upper bound for this is $n^2 T(k, m)$. As seen in the previous section, a matching can be computed in $O(mn + m\sqrt{m})$ time [7]. So, we spend $O(n^2 T(k,m) + mn + m\sqrt{m})$ time in total on the decision algorithm.

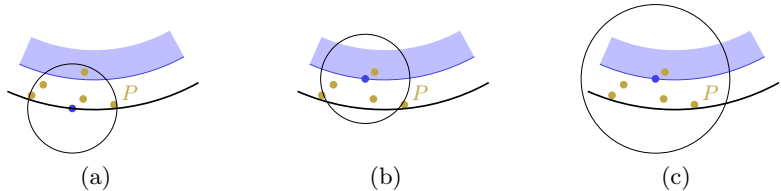

**Fig. 14.** (a) A circle of radius $cd$ covers a number of points of $P_I$ inside a certain cell $\mathcal{A}_I$ of the arrangement. The centre $q$ of the circle lies inside $\mathcal{A}_I$, but not inside the region $\tilde{q}$. (b) The point $q$ has been moved into $\tilde{q}$, but now some points of $P_I$ that were covered are no longer covered. (c) The enlarged circle of radius $(c+2)d$ covers the points again.